\documentclass[conference]{IEEEtran}

\usepackage[colorlinks]{hyperref}
\usepackage{makeidx}
\makeindex

\usepackage{amsmath}
\usepackage{amsthm}
\usepackage{amssymb}
\usepackage{bm}
\usepackage{xspace}
\usepackage{xcolor}
\usepackage{graphicx}
\usepackage{url}
\usepackage{framed}
\usepackage{float}
\usepackage{rotating}
\usepackage{verbatim}
\usepackage{listings}
\usepackage{lscape}

\usepackage{pgfplots}
\usepackage{pgf}
\usepackage{tikz}
\usetikzlibrary{arrows,shapes.misc,chains,scopes}
\pgfplotsset{compat = newest}
\usepackage{pgfplotstable}
\usepackage{booktabs}
\usepackage{colortbl}

\newcommand{\executeiffilenewer}[3]{%
\ifnum\pdfstrcmp{\pdffilemoddate{#1}}%
{\pdffilemoddate{#2}}>0%
{\immediate\write18{#3}}\fi%
}

\graphicspath{{figures/}}

\theoremstyle{plain}
\newtheorem{proposition}{Proposition}

\newcounter{algocount}
\newcounter{examplecount}

\newcommand{\setz}{\ensuremath{\mathbf{Z}}\xspace}

\newcommand{\setb}{\ensuremath{\mathcal{B}}\xspace}

\newcommand{\setd}{\ensuremath{\mathcal{D}}\xspace}
\newcommand{\setl}{\ensuremath{\mathcal{L}}\xspace}

\newcommand{\sett}{\ensuremath{\mathcal{T}}\xspace}

\newcommand{\setx}{\ensuremath{\mathcal{X}}\xspace}

\DeclareMathOperator{\expop}{\mathbb{E}}
\DeclareMathOperator{\entop}{\mathbb{H}}

\DeclareMathOperator{\kl}{\mathbb{D}}

\usepackage{cite}
\usepackage{printlen}
\usepackage{booktabs}

\newcommand{\oleq}[1]{\overset{\text{(#1)}}{\leq}}
\newcommand{\ogeq}[1]{\overset{\text{(#1)}}{\geq}}
\newcommand{\oeq}[1]{\overset{(\text{#1})}{=}}

\newcommand{\bsm}[1]{\begin{array}{#1}}
\newcommand{\esm}{\end{array}}

\renewcommand{\setz}{\mathcal{Z}}

\definecolor{bblue}{rgb}{0.2,0.2,0.7}

\title{Informational Divergence and Entropy Rate\\on Rooted Trees with Probabilities\\[-0.0cm]}

\IEEEoverridecommandlockouts

\author{\IEEEauthorblockN{Georg B\"ocherer and Rana Ali Amjad}
\IEEEauthorblockA{Institute for Communications Engineering\\Technische Universit\"at M\"unchen, Germany\\
Email: \texttt{georg.boecherer@tum.de,raa2463@gmail.com}}
}

\DeclareMathOperator{\supp}{supp}

\begin{document}
\maketitle

\begin{abstract}
Rooted trees with probabilities are used to analyze properties of a variable length code. A bound is derived on the difference between the entropy rates of the code and a memoryless source. The bound is in terms of normalized informational divergence. The bound is used to derive converses for exact random number generation, resolution coding, and distribution matching.
\end{abstract}
\section{Introduction}
A rooted tree with probabilities is shown in Fig.~\ref{fig:tree}. The tree consists of a root $\epsilon$, branching nodes $\{\epsilon,1\}$, and leaves $\{0,10,11\}$. $P_Y$ is the leaf distribution. James L. Massey advocated the framework of such trees for the analysis of variable length codes \cite{massey1983entropy},\cite{rueppel1994leaf},\cite[Sec.~2.2.2]{masseyapplied1}. 

Consider a discrete memoryless source (DMS) $P_Z$ with letters in $\setz$ and consider a device that generates variable length codewords with letters in $\setz$. We are interested in two properties.
\begin{itemize}
\item[(1)] How well does our device mimic the DMS $P_Z$?
\item[(2)] At which rate does our device produce output?
\end{itemize}
We measure (1) by \emph{normalized informational divergence} and (2) by \emph{entropy rate}. In this work, we use the framework of rooted trees with probabilities to relate these two measures.

This paper is organized as follows. In Sec.~\ref{sec:rooted}, we review properties of rooted trees with probabilities. In Sec.~\ref{sec:chain}, we derive chain rules for such trees by using Rueppel and Massey's Leaf-Average Node Sum Interchange Theorem (LANSIT) \cite{rueppel1994leaf}. We propose a \emph{normalized} LANSIT and state \emph{normalized} chain rules. In Sec~\ref{sec:divent}, we derive variable length results for normalized informational divergence and entropy rate. In Sec.~\ref{sec:converses}, we apply our results to derive converses, which recover existing converses for exact random number generation \cite{knuth1976complexity,han1997interval} and generalize existing converses for resolution coding \cite[Sec.~II]{han1993approximation},\cite{bocherer2013fixed}. We establish a new converse for distribution matching \cite{bocherer2011matching,bocherer2012capacity,amjad2013fixed}.
\section{Rooted Trees with Probabilities}
\begin{figure}
\centering
\scalebox{1.0}{		\tikzstyle{branching}=[circle,draw=black!50,fill=black!20,thick,inner sep=0pt,minimum size=5mm]
		\tikzstyle{leaf}=[circle,draw=green!50!black,fill=green!20,thick,inner sep=0pt,minimum size=6mm]
		\tikzstyle{emptyleaf}=[circle,draw=red!50,fill=red!20,thick,inner sep=0pt,minimum size=6mm]
		\tikzstyle{label0} = [above]
		\tikzstyle{label1} = [below]

\begin{tikzpicture}[node distance=0cm, grow=right, scale=0.9,level 1/.style={sibling distance=2.1cm,level distance=1.5cm},level 2/.style={sibling distance=1.4cm,level distance=1.5cm},level 3/.style={sibling distance=1cm,level distance=1cm},unit/.style={rectangle,fill=black!5,draw,thick,minimum size = 0.75cm, text width = 1.5cm, text centered}]
\node[branching] at (0,0) {$\epsilon$}
	child {node[branching] {$1$}
		child {node[leaf](11) {$11$}
			edge from parent[->]
			node[label1]{$1$}
		}
		child {node[leaf](10) {$10$}
			edge from parent[->]
			node[label0]{$0$}
		}
	edge from parent[->]
	node[label1]{$1$}
	}
	child {node[leaf](0) {$0$}
	edge from parent[->]
	node[label0]{$0$}
	};
\node [right=of 0]{$P_Y(0)=\frac{1}{2}$};
\node [right=of 10]{$P_Y(10)=\frac{1}{8}$};
\node [right=of 11]{$P_Y(11)=\frac{3}{8}$};






\end{tikzpicture}}
\caption{A rooted tree with probabilities over the binary alphabet $\setz=\{0,1\}$.}
\label{fig:tree}
\vspace{-0.5cm}
\end{figure}
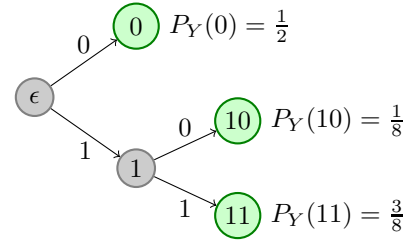
\label{sec:rooted}
\begin{table*}
\caption{Chain Rules on Rooted Trees with Probabilities. }
\label{tab:chain}
\centering
\begin{tabular}{ccccc}
&function&increment&un-normalized&normalized\\\hline
\addlinespace
LANSIT&$f(t)$&$\Delta f(tz)$&$\displaystyle\expop[f(Y)]-f(\varepsilon)=\sum_{t\in\mathcal{B}}Q(t)\expop[\Delta f(tY_t)]$&$\frac{\expop[f(Y)]-f(\varepsilon)}{\expop[\ell(Y)]}=\displaystyle\sum_{t\in\setb}P_B(t)\expop[\Delta f(tY_t)]$
\\\hline
\addlinespace
Path Length Lemma&$\ell(t)$&$1$&$\displaystyle\expop[\ell(Y)]=\sum_{t\in\mathcal{B}}Q(t)$&$\frac{\expop[\ell(Y)]}{\expop[\ell(Y)]}=\displaystyle\sum_{t\in\setb}P_B(t)\expop[\Delta \ell(tY_t)]=1$\\
\addlinespace
Leaf Entropy Lemma&$-\log_2 Q(t)$&$-\log_2 P_{Y_t}(z)$&$\displaystyle\entop(P_Y)=\sum_{t\in\mathcal{B}}Q(t)\entop(P_{Y_t})$&$\frac{\entop(P_Y)}{\expop[\ell(Y)]}=\entop(P_{Y_B}|P_B)$\\
\addlinespace
Leaf Divergence Lemma&$\log_2 \frac{Q(t)}{Q'(t)}$&$\log_2 \frac{P_{Y_t}(z)}{P_{Y'_t}(z)}$&$\displaystyle\kl(P_Y\Vert P_{Y'})=\sum_{t\in\mathcal{B}}Q(t)\kl(P_{Y_t}\Vert P_{Y'_t})$&$\frac{\kl(P_Y\Vert P_{Y'})}{\expop[\ell(Y)]}=\kl(P_{Y_B}\Vert P_{Y'_B}|P_B)$\\\hline
\end{tabular}
\vspace{-0.5cm}
\end{table*}
We consider finite rooted trees over finite alphabets $\mathcal{Z}=\{0,1,\dotsc,m-1\}$. An example for $\setz=\{0,1\}$ is shown in Fig.~\ref{fig:tree}. 
A rooted tree $\mathcal{T}$ consists of branching nodes $\mathcal{B}$ with $m$ successors each and leaves $\mathcal{L}$ with no successors. Each node except the root node has exactly one predecessor. The root node has no predecessor. For each branching node, each element of $\setz$ labels exactly one outgoing branch. Each node is uniquely identified by the string of labels on the path from the root to the node. The root node is identified by the empty string $\varepsilon$.
For each node $t\in\mathcal{T}$, $\ell(t)$ denotes the number of branches on the path from the root to the node $t$. Equivalently, $\ell(t)$ is the number of letters in the string $t$.
\subsection{Probabilities Induced by Leaf Distribution}
Consider a random variable $Y$ with distribution $P_Y$ on $\mathcal{L}$. We extend $P_Y$ to $\sett$ by associating with each string $t\in\mathcal{T}$ a probability
\begin{align}
Q(t)=\sum_{s\in\mathcal{L}\colon s_1^{\ell(t)}=t}P_Y(s)
\end{align}
where $s_1^{\ell(t)}=s_1s_2\dotsb s_{\ell(t)}$. In other words, $Q(t)$ is the sum of the probabilities of all leaves that have $t$ as a prefix. We can interpret $Q(t)$ as the probability of choosing a path from the root to a leaf that passes through node $t$. In particular, the node probability of the root is always $Q(\epsilon)=1$. For example, the node probabilities in Fig.~\ref{fig:tree} are
\begin{align}
Q(\epsilon)=1,\;\;Q(0)=Q(1)=\frac{1}{2},\;\;Q(10)=\frac{1}{8},\;\;Q(11)=\frac{3}{8}.\nonumber 
\end{align}
For each string $t\in\setb$ and each letter $z\in\setz$, we define a branching probability
\begin{align}
P_{Y_t}(z)=\frac{Q(tz)}{Q(t)}
\end{align}
where $tz$ is the string $t$ concatenated with the letter $z\in\setz$. The branching distributions in Fig.~\ref{fig:tree} are thus
\begin{align}
&P_{Y_1}(0)=\frac{\frac{1}{8}}{\frac{1}{2}}=\frac{1}{4},\quad P_{Y_1}(1)=\frac{\frac{3}{8}}{\frac{1}{2}}=\frac{3}{4},\;
\\
&P_{Y_\epsilon}(0)=P_{Y_\epsilon}(1)=\frac{1}{2}.\label{eq:memory}
\end{align}
\subsection{Probabilities Induced by Alphabet Distribution}
Let $P_Z$ be a distribution on the alphabet $\setz$. The distribution $P_Z$ induces a distribution on $\setl$, which we denote by $P_Z^\setl$. For each $t\in\setl$, we have
\begin{align}
P_Z^\setl(t):=P_Z(t_1)\dotsb P_Z(t_{\ell(t)}).
\end{align}
For example, consider the binary distribution $P_Z(0)=1-P_Z(1)=\frac{1}{3}$. For the leaves $\setl=\{0,10,11\}$ in Fig.~\ref{fig:tree}, the distribution $P_Z$ induces the distribution
\begin{align}
P_Z^\setl(0)=\frac{1}{3},\quad P_Z^\setl(10)=\frac{2}{3}\cdot\frac{1}{3},\quad P_Z^\setl(11)=\frac{2}{3}\cdot\frac{2}{3}.\label{eq:pzc}
\end{align}
If all strings in $\setl$ are of length $n$, then $P_Z^\setl(t)=P_Z^n(t)$ for all $t\in\setl$, where $P_Z^n$ is the usual product distribution of $n$ independent random variables with distribution $P_Z$.

\section{Chain Rules on Trees}
\label{sec:chain}
\subsection{Notation}
We denote expectation by $\expop[\cdot]$ and define informational divergence, entropy, and variational distance as
\begin{align}
&\kl(P_Y\Vert P_Z):=\sum_{z\in\supp P_Y}P_Y(z)\log_2\frac{P_Y(z)}{P_Z(z)}\\
&\entop(P_Y):=\sum_{z\in\supp P_Y}P_Y(z)[-\log_2 P_Y(z)]\\
&\lVert P_Y-P_Z\rVert_1:=\sum_{z\in\supp P_Y\cup\supp P_Z}|P_Y(z)-P_Z(z)|
\end{align}
where $\supp P_Y$ is the support of $P_Y$. 
\subsection{LANSIT}
Let $\mathcal{T}$ be a rooted tree and let $f$ be a function that assigns to each $t\in\mathcal{T}$ a real value $f(t)$. For each $t\in\setb$ and $z\in\setz$ define the increment $\Delta f(tz):=f(tz)-f(t)$. Rueppel and Massey's LANSIT is the following general chain rule. 
\begin{proposition}[LANSIT, \text{\cite[Theo~1]{rueppel1994leaf}}]\label{prop:integrallansit}
\begin{align}
\expop[f(Y)]-f(\varepsilon)=\sum_{t\in\mathcal{B}}Q(t)\expop[\Delta f(tY_t)].\label{eq:integrallansit}
\end{align}
\end{proposition}
In Tab.~\ref{tab:chain}, we display various instances of the LANSIT. The Path Length Lemma and the Leaf Entropy Lemma can be found, e.g, in Massey's lecture notes \cite[Sec.~2.2.2]{masseyapplied1}. The Leaf Divergence Lemma is to the best of our knowledge stated here for the first time. If all paths in a tree have the same length $n$,  then $P_Y=P_{Y^n}$ is a joint distribution of a random vector $Y^n=Y_1Y_2\dotsb Y_n$ that takes on values in $\setz^n$. For $i=\ell(t)$, we have $P_{Y_t}=P_{Y_{i+1}|Y_1^i}(\cdot|t)$ and $Q(t)=P_{Y_1^i}(t)$ and the Leaf Entropy Lemma and the Leaf Divergence Lemma are the usual chain rules for entropy and informational divergence, respectively \cite[Chap.~2]{cover2006elements}.
\subsection{Normalized LANSIT}
Let $B$ be a random variable on the set of branching nodes $\setb$ and define
\begin{align}
P_B(t)=\frac{Q(t)}{\expop[\ell(Y)]},\quad t\in\setb.\label{eq:defpb}
\end{align}
We have
\begin{align}
\sum_{t\in\mathcal{B}}P_B(t)\oeq{a}\frac{\sum_{t\in\mathcal{B}}Q(t)}{\expop[\ell(Y)]}\oeq{b}1\label{eq:pbdist}
\end{align}
where (a) follows by the definition of $P_B$, and (b) follows by the Path Length Lemma. It follows from \eqref{eq:pbdist} that $P_B$ defines a distribution on $\mathcal{B}$. This observation leads to the following simple and useful extension of the LANSIT. 
\begin{proposition}[Normalized LANSIT]
\begin{align}
\frac{\expop[f(Y)]-f(\varepsilon)}{\expop[\ell(Y)]}=\sum_{t\in\setb}P_B(t)\expop[\Delta f(tY_t)].\label{eq:differentiallansit}
\end{align}
\end{proposition}
For a real-valued function $g$ defined on the set of distributions, we use the notation
\begin{align}
\expop[g(P_{Y_B})|P_B]:=\sum_{t\in\setb}P_B(t)g(P_{Y_t}).\label{eq:conditionaldef}
\end{align}
Accordingly, we define $\entop(P_{Y_B}|P_B)$ and $\kl(P_{Y_B}\Vert P_{Y'_B}|P_B)$. Using this notation, we list normalized versions of the Path Length Lemma, the Leaf Entropy Lemma, and the Leaf Divergence Lemma in Tab.~\ref{tab:chain}. These normalized versions are instances of the normalized LANSIT.
\section{Informational Divergence and Entropy Rate}
\label{sec:divent}
We compare an arbitrary distribution $P_Y$ on the set of leaves $\mathcal{L}$ to the distribution $P_Z^\mathcal{L}$ on $\mathcal{L}$ that is induced by a DMS $P_Z$. Note that in general, $P_Y$ generates letters from $\mathcal{Z}$ with memory, see Fig.~\ref{fig:tree} and \eqref{eq:memory} for an example. 
\subsection{Codewords of Length $1$}
We start with the special case when $\ell(t)=1$ for all $t\in\setl$ and equivalently, $\setl=\setz$. The DMS we compare to is the uniform distribution $P_U$ on $\mathcal{Z}$. In this case, normalized and un-normalized informational divergence are the same and entropy rate is the same as entropy. We have
\begin{align}
\kl(P_Y\Vert P_U)=\entop(P_U)-\entop(P_Y).\label{eq:uniformRelation}
\end{align}
In particular, if $\kl(P_Y\Vert P_U)\to 0$ then $\entop(P_Y)\to \entop(P_U)$. Next, suppose the DMS we compare to has a distribution $P_Z$ that is not necessarily uniform. By Pinsker's inequality \cite[Lemma~11.6.1]{cover2006elements}, we have
\begin{align}
\kl(P_Y\Vert P_Z)\to 0\Rightarrow \lVert P_Y-P_Z\rVert_1\to 0.\label{eq:firstImplication}
\end{align}
Let $g$ be a function that is continuous in $P_Z$. Then we have
\begin{align}
\lVert P_Y-P_Z\rVert_1\to 0\Rightarrow |g(P_Y)-g(P_Z)|\to 0.\label{eq:secondImplication}
\end{align}
For instance, the entropy $\entop$ is continuous in $P_Z$ \cite[Lemma~2.7]{csiszar2011information} and therefore
\begin{align}
\lVert P_Y-P_Z\rVert_1\to 0\Rightarrow |\entop(P_Y)-\entop(P_Z)|\to 0.\label{eq:thirdImplication}
\end{align}
Combining \eqref{eq:firstImplication} and \eqref{eq:thirdImplication}, we get the relation
\begin{align}
\kl(P_Y\Vert P_Z)\to  0\Rightarrow |\entop(P_Y)-\entop(P_Z)|\to 0.\label{eq:lastImplication}
\end{align}

\subsection{Codewords of Length Larger than $1$: First Attempt}

Consider the special case when the generated strings are of fixed length $n\geq 1$ with the joint distribution $P_{Y^n}$. Suppose further that
\begin{align}
\frac{\kl(P_{Y^n}\Vert P_Z^n)}{n}\leq \frac{\sqrt{n}}{n}.
\end{align}
As $n\to\infty$, the normalized informational divergence approaches zero. By Pinsker's inequality, we have
\begin{align}
\lVert P_{Y^n}-P_Z^n\rVert_1\leq\sqrt{\sqrt{n}2\ln 2}.\label{eq:failure}
\end{align}
For $n\geq 9$, the right-hand side of \eqref{eq:failure} is larger than $2$, which is useless because variational distance is trivially bounded from above by $2$. This example illustrates that the line of arguments \eqref{eq:firstImplication}--\eqref{eq:lastImplication} does not directly generalize to codeword lengths larger than one. This is our motivation to analyze the variable length case within the framework of rooted trees.

\subsection{Normalized Pinsker's Inequality}

\begin{proposition}[Normalized Pinsker's Inequality]\label{prop:pinskertree}
\begin{align}
\frac{\kl(P_Y\Vert P_Z^\mathcal{L})}{\expop[\ell(Y)]} \oeq{a} &\kl(P_{Y_B}\Vert P_Z|P_B)\nonumber\\
\ogeq{b}&\frac{1}{2\ln 2}\expop\Bigl[\lVert P_{Y_B}-P_Z\rVert_1^2\Big|P_B\Bigr]\nonumber\\
\ogeq{c}&\frac{1}{2\ln 2}\expop^2\Bigl[\lVert P_{Y_B}-P_Z\rVert_1\Big|P_B\Bigr].
\label{eq:pinskertree}
\end{align}
\end{proposition}
\begin{IEEEproof}
Equality in (a) follows by the Normalized Leaf Divergence Lemma, (b) follows by Pinsker's inequality, and (c) follows by Jensen's inequality \cite[Chap.~2]{cover2006elements}.
\end{IEEEproof}
Prop.~\ref{prop:pinskertree} is a quantitative statement. Qualitatively, we have
\begin{align}
\frac{\kl(P_Y\Vert P_Z^\setl)}{\expop[\ell(Y)]}\to 0\Rightarrow \expop\Bigl[\lVert P_{Y_B}-P_Z\rVert_1\Big|P_B\Bigr]\to 0.\label{eq:firstImplicationN}
\end{align}
If $\mathcal{L}=\mathcal{Z}$, i.e., all strings in $\mathcal{L}$ are of length $1$ and $\setb=\{\varepsilon\}$, then \eqref{eq:pinskertree} is simply the original Pinsker's inequality and  \eqref{eq:firstImplicationN} recovers implication \eqref{eq:firstImplication}.
\subsection{Continuity for Trees}
\begin{proposition}\label{prop:continuity}
Let $P$ be a distribution on $\mathcal{Z}$ and let $g$ be a real-valued function whose maximum and minimum values differ at most by $g_{\max}$. Suppose that $g$ is continuous in $P_Z$, i.e., there is a function $\delta(\epsilon)$ such that for all $\epsilon\geq 0$ 
\begin{align}
\lVert P-P_Z\rVert_1\leq\epsilon\Rightarrow|g(P)-g(P_Z)|\leq\delta(\epsilon)\label{eq:continuity}
\end{align}
where $\delta(\epsilon)\to 0$ as $\epsilon\to 0$. Then we have for all $\epsilon\geq 0$
\begin{align}
\expop&\Bigl[\lVert P_{Y_B}-P_Z\rVert_1\Big|P_B\Bigr]\leq\theta\nonumber\\
&\Rightarrow\;\Bigl|\expop\bigl[g(P_{Y_B})\big|P_B\bigr]-g(P_Z)\Bigr|\leq\delta(\epsilon)+\frac{\theta}{\epsilon}g_{\max}.\label{eq:continuityg}
\end{align}
\end{proposition}
\begin{IEEEproof}
The proof is given in Appendix~\ref{proof:continuity}.
\end{IEEEproof}
By setting $\theta=\epsilon^2$ in \eqref{eq:continuityg}, we get the qualitative implication
\begin{align}
\expop&\Bigl[\lVert P_{Y_B}-P_Z\rVert_1\Big|P_B\Bigr]\to 0\nonumber\\
&\Rightarrow\Bigl|\expop\bigl[g(P_{Y_B})\big|P_B\bigr]-g(P_Z)\Bigr|\to 0.\label{eq:secondImplicationN}
\end{align}
If $\mathcal{L}=\setz$, then \eqref{eq:secondImplicationN} recovers \eqref{eq:secondImplication} for bounded $g$. For a specific function $g$, if the function $\delta(\epsilon)$ is known, then the right-hand side of \eqref{eq:continuityg} can be minimized over $\epsilon$ to get a bound that depends only on $\theta$. 
\subsection{Entropy Rate Continuity on Trees}
The entropy is continuous in $P_Z$ \cite[Lemma~2.7]{csiszar2011information} and bounded by $\log_2|\setz|$. Thus, Prop.~\ref{prop:continuity} applies for $g=\entop$ and we have the implication
\begin{align}
\expop\Bigl[\lVert P_{Y_B}-P_Z\rVert_1\Big|P_B\Bigr]\to 0&\nonumber\\
\Rightarrow\left|\frac{\entop(P_Y)}{\expop[\ell(Y)]}-\entop(P_Z)\right|&\oeq{a}\Bigl|\entop(P_{Y_B}|P_B)-\entop(P_Z)\Bigr|\nonumber\\
&\overset{\text{(b)}}{\to} 0.\label{eq:thirdImplicationN}
\end{align}
Step (a) follows by the Normalized Leaf Entropy Lemma and (b) follows by \eqref{eq:secondImplicationN}. Note that by the Normalized Leaf Divergence Lemma, $\entop(P_Z)$ is the entropy rate of $P_Z^\setl$. For $\mathcal{L}=\setz$, \eqref{eq:thirdImplicationN} recovers implication \eqref{eq:thirdImplication}.
\begin{proposition}\label{prop:entropyrate}
The following implication holds:
\begin{align}
\expop&\Bigl[\lVert P_{Y_B}-P_Z\rVert_1\Big|P_B\Bigr]\leq\theta(\epsilon)\nonumber\\
&\Rightarrow\left|\frac{\entop(P_Y)}{\expop[\ell(Y)]}-\entop(P_Z)\right|\leq\sigma(\epsilon),\quad 0\leq\epsilon\leq\frac{1}{2}\label{eq:rhosigma}
\end{align}
where
\begin{align}
\theta(\epsilon)=&\frac{1}{\log_2|\setz|}\epsilon^2\log_2\frac{|\setz|}{e\epsilon},\quad
\sigma(\epsilon)=\epsilon\log_2\frac{|\setz|^2}{e\epsilon^2}.\label{eq:rhosigma2}
\end{align}
\end{proposition}
\begin{IEEEproof}
The proof is given in Appendix~\ref{proof:entropyrate}.
\end{IEEEproof}
In \eqref{eq:rhosigma}, as $\epsilon\to 0$, both $\theta(\epsilon)\to 0$ and $\sigma(\epsilon)\to 0$, which shows that Prop.~\ref{prop:entropyrate} provides a quantitative version of \eqref{eq:thirdImplicationN}.
\subsection{Normalized Informational Divergence and Entropy Rate}
By the qualitative implications \eqref{eq:firstImplicationN} and \eqref{eq:thirdImplicationN}, we have
\begin{align}
\frac{\kl(P_Y\Vert P_Z^\mathcal{L})}{\expop[\ell(Y)]}\to 0\Rightarrow\left|\frac{\entop(P_Y)}{\expop[\ell(Y)]}-\entop(P_Z)\right|\to 0.\label{eq:fourthImplicationN}
\end{align}
For $\mathcal{L}=\setz$, \eqref{eq:fourthImplicationN} recovers implication \eqref{eq:lastImplication}. The next proposition provides a quantitative version of implication \eqref{eq:fourthImplicationN}.
\begin{proposition}\label{prop:implication4}
Let $\theta,\sigma$ be the functions defined in Prop.~\ref{prop:entropyrate}. For $0\leq\alpha\leq\frac{1}{2\ln 2}\theta^2(\frac{1}{2})$, define $\epsilon'=\theta^{-1}(\sqrt{\alpha 2\ln 2})$ and $\beta(\alpha)=\sigma(\epsilon')$. We have the implication
\begin{align}
\frac{\kl(P_Y\Vert P_Z^\mathcal{L})}{\expop[\ell(Y)]}\leq\alpha
\Rightarrow&\left|\frac{\entop(P_Y)}{\expop[\ell(Y)]}-\entop(P_Z)\right|\leq\beta(\alpha)\label{eq:alphabeta}\\
\alpha\to 0\Rightarrow &\beta(\alpha)\to 0.\label{eq:alphabeta2}
\end{align}
\end{proposition}
\begin{IEEEproof}
Statement \eqref{eq:alphabeta} follows by combining Prop.~\ref{prop:pinskertree} and Prop.~\ref{prop:entropyrate}.
As $\alpha\to 0$, $\epsilon'=\theta^{-1}(\sqrt{\alpha 2\ln 2})\to 0$ by \eqref{eq:rhosigma2} and therefore, $\beta(\alpha)=\sigma(\epsilon')\to 0$. This proves \eqref{eq:alphabeta2}.
\end{IEEEproof}
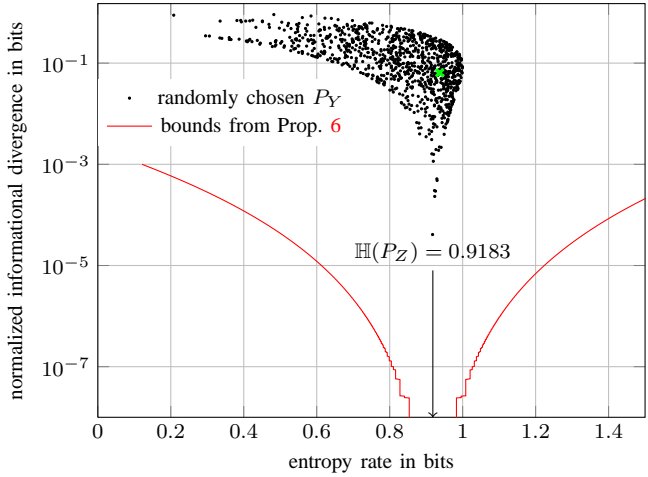
\begin{figure}
\footnotesize
\pgfplotsset{
width=\columnwidth,
height=0.3\textheight
}
\pgfplotsset{every axis title/.append style={at={(0.0,1.05)},
right
}}
\begin{tikzpicture}
\begin{semilogyaxis}[
xmin=0,
xmax=1.5,
ymin=0.00000001,
ymax=1.5,
xlabel={entropy rate in bits},
ylabel={normalized informational divergence in bits},
grid=both,
legend columns=1,
legend style={at={(0.0,0.82)},anchor=north west,draw=none},
legend entries = {randomly chosen $P_Y$,bounds from Prop.~\ref{prop:implication4}},
]
\addplot[black,only marks,mark options={scale=0.2}]
table[x=ent,y=kl]
{examplepoints.txt};
\addplot[red,no markers]
table[x=entm,y=kl]
{examplebounds.txt};
\addplot[red,no markers]
table[x=entp,y=kl]
{examplebounds.txt};

\addplot[green,only marks, mark=x,line width=1.2pt]
coordinates{(0.9371,0.0645)}
;

\draw[<-] (axis cs:0.9183,0.00000001) node {} -- (axis cs:0.9183,0.000008) node[anchor=south]{$\entop(P_Z)=0.9183$};

\end{semilogyaxis}
\end{tikzpicture}
\caption{Normalized informational divergence versus entropy rate for the rooted tree in Fig.~\ref{fig:tree}. We compare a leaf distribution $P_Y$ with the DMS $P_Z(0)=1-P_Z(1)=\frac{1}{3}$. The induced leaf distribution $P_Z^\setl$ is calculated in \eqref{eq:pzc}. In horizontal direction, we display $\entop(P_Y)/\expop[\ell(Y)]$ and in vertical direction $\kl(P_Y\Vert P_Z^\setl)/\expop[\ell(Y)]$. The green cross corresponds to the $P_Y$ stated in Fig.~\ref{fig:tree}. The black dots result from $1000$ distributions $P_Y$ that were generated by choosing the entries uniformly at random between zero and one and then normalizing to one. The black arrow indicates the point that corresponds to $P_Y=P_Z^\setl$. The red lines display the bounds from Prop.~\ref{prop:implication4} with $\alpha$ in vertical and $\entop(P_Z) \pm \beta(\alpha)$ in horizontal direction. Note that the red bounds apply to \emph{any} rooted tree with alphabet size $|\setz|=2$. Changing $P_Z$ and thereby $\entop(P_Z)$ changes only the horizontal position of the bounds.}
\label{fig:bounds}
\end{figure}
In Fig.~\ref{fig:bounds}, we display the bounds from Prop.~\ref{prop:implication4} for the rooted tree in Fig.~\ref{fig:tree}.
%
%
%
%
\section{Converses}
\label{sec:converses}
We want to encode a DMS $P_X$ with letters in $\setx$ to mimic a target DMS $P_Z$ with letters in $\setz$. Variable length coding uses a complete dictionary $\mathcal{D}$ with letters in $\mathcal{X}$, a complete codebook $\mathcal{C}$ with letters in $\mathcal{Z}$, and a mapping $f\colon\mathcal{D}\to\mathcal{C}$. A set is complete if it is the set of leaves of a rooted tree as defined in Sec.~\ref{sec:rooted}. The encoder parses the input stream by the dictionary, which generates a random variable $D$ with distribution $P_X^\setd$. The mapping generates a random variable $Y=f(D)$.  Two classes of mappings are of interest.
\begin{enumerate}
\item[1.] The mapping $f$ is \emph{deterministic} but the input does not need to be reconstructed from the output.
\item[2.] The mapping $f$ is \emph{random} but the input has to be reconstructed correctly from the output with probability close to one. 
\end{enumerate}
In the following, we derive rate converses for encoders in class 1. (2.) that bound the minimum (maximum) rate, at which a required normalized informational divergence can be achieved.

\subsection{Converse for Deterministic Encoders}
Consider an encoder of class 1. Since the mapping $f$ is deterministic, we have
\begin{align}
\entop(P_D)=\entop(P_{Df(D)})\geq\entop(P_{f(D)})=\entop(P_Y).\label{eq:detbound}
\end{align}
By the Normalized Leaf Entropy Lemma, we have
\begin{align}
\entop(P_D)=\entop(P_X^\setd)=\expop[\ell(D)]\entop(P_X).\label{eq:nlel}
\end{align}
Suppose $\kl(P_Y\Vert P_Z^\mathcal{C})/\expop[\ell(Y)]\leq\alpha$. Then by Prop~\ref{prop:implication4} we have
\begin{align}
\entop(P_Y)\geq \entop(P_Z)\expop[\ell(Y)]-\beta(\alpha)\expop[\ell(Y)].\label{eq:detdivent}
\end{align}
Using \eqref{eq:nlel} and \eqref{eq:detdivent} in \eqref{eq:detbound} and reordering the terms gives the following result.
\begin{proposition}\label{prop:rngconverse}
\begin{align}
\frac{\kl(P_Y\Vert P_Z^\mathcal{C})}{\expop[\ell(Y)]}\leq\alpha\Rightarrow\frac{\expop[\ell(D)]}{\expop[\ell(Y)]}\geq\frac{\entop(P_Z)}{\entop(P_X)}-\frac{\beta(\alpha)}{\entop(P_X)}.
\end{align}
\end{proposition}
Since $\expop[\ell(Y)]\geq 1$, Prop.~\ref{prop:rngconverse} provides a rate converse also for un-normalized informational divergence. Prop.~\ref{prop:rngconverse} establishes quantitative variable-length versions of the converses in \cite[Sec.~II]{han1993approximation} both for normalized and un-normalized informational divergence. Prop.~\ref{prop:rngconverse} implies \cite[Prop.~III]{bocherer2013fixed}. Exact generation of $P_Z$ requires $\alpha=0$, which implies $\beta(\alpha)=0$, and we recover the converses by Knuth and Yao \cite{knuth1976complexity} and Han and Hoshi \cite{han1997interval}. 

\subsection{Converse for Random Encoders}

Let $g$ be a decoder that calculates an estimate $\hat{D}=g(Y)$ and let $P_e:=\Pr\{\hat{D}\neq D\}$ be the probability of erroneous decoding. We have
\begin{align}
\entop(P_D)-\entop(P_Y)&\leq \entop(P_D|P_Y)\nonumber\\
&\oleq{a}\entop_2(P_e)+P_e\log_2|\setd|\label{eq:decbound}
\end{align}
where (a) follows by Fano's inequality \cite[Theo.~2.10.1]{cover2006elements} and where $\entop_2$ denotes the binary entropy function. Suppose $\kl(P_Y\Vert P_Z^\mathcal{C})/\expop[\ell(Y)]\leq\alpha$. Then by Prop~\ref{prop:implication4} we have
\begin{align}
\entop(P_Y)\leq \entop(P_Z)\expop[\ell(Y)]+\beta(\alpha)\expop[\ell(Y)].\label{eq:decdivent}
\end{align}
Combining \eqref{eq:decbound}, \eqref{eq:decdivent}, and \eqref{eq:nlel} and reordering the terms proves the following proposition.
\begin{proposition}\label{eq:decconverse}
The inequalities
\begin{align}
P_e\leq\epsilon\leq \frac{1}{2},\quad\frac{\kl(P_Y\Vert P_Z^\mathcal{C})}{\expop[\ell(Y)]}\leq\alpha
\end{align}
imply
\begin{align}
\frac{\expop[\ell(D)]}{\expop[\ell(Y)]}\leq \frac{\entop(Z)}{\entop(X)}+\frac{\beta(\alpha)}{\entop(X)}+\frac{\entop_2(\epsilon)+\epsilon\log_2|\setd|}{\expop[\ell(Y)]\entop(X)}.\label{eq:randomconverse}
\end{align}
\end{proposition}
Inequality \eqref{eq:randomconverse} establishes a rate converse for distribution matching. Variable length codes for which achievability can be shown are presented in \cite{bocherer2011matching},\cite{bocherer2012capacity},\cite{amjad2013fixed}.

\appendix

\section{Proofs}
\subsection{Proof of Prop.~\ref{prop:continuity}}
\label{proof:continuity}
We have
\begin{align} 	
&\hspace{-0.5cm}|\expop[g(P_{Y_B})|P_B]-g(P_Z)|=\Bigl|\sum_{t\in\mathcal{B}} P_B(t)[g(P_{Y_t})-g(P_Z)]\Bigr|\nonumber\\
\leq&\sum_{t\in\setb} P_B(t)\bigl|g(P_{Y_t})-g(P_Z)\bigr|\\
=&\sum_{t\colon \lVert P_{Y_t}-P_Z\rVert_1<\epsilon}P_B(t)\bigl|g(P_{Y_t})-g(P_Z)\bigr|\nonumber\\
&\qquad+\sum_{t\colon \lVert P_{Y_t}-P_Z\rVert_1\geq\epsilon}P_B(t)\bigl|g(P_{Y_t})-g(P_Z)\bigr|.\label{eq:iii:sums}
\end{align}
We next bound the two sums in \eqref{eq:iii:sums}. The first sum in \eqref{eq:iii:sums} is bounded by 
\begin{align}
\sum_{t\colon \lVert P_{Y_t}-P_Z\rVert_1<\epsilon}&P_B(t)\bigl|g(P_{Y_t})-g(P_Z)\bigr|\nonumber\\
&\oleq{a}\sum_{t\colon \lVert P_{Y_t}-P_Z\rVert_1<\epsilon}P_B(t)\delta(\epsilon)\nonumber\\
&\leq\delta(\epsilon)\label{eq:sum1bound}
\end{align}
where (a) follows from \eqref{eq:continuity}. The second sum in \eqref{eq:iii:sums} is bounded as
\begin{align} 	
\sum_{t\colon \lVert P_{Y_t}-P_Z\rVert_1\geq\epsilon}&P_B(t)\bigl|g(P_{Y_t})-g(P_Z)\bigr|\nonumber\\
&\oleq{a}\sum_{t\colon \lVert P_{Y_t}-P_Z\rVert_1\geq\epsilon}P_B(t) g_{\max}\nonumber\\
&\leq g_{\max}\sum_{t\colon \lVert P_{Y_t}-P_Z\rVert_1\geq\epsilon}P_B(t)\frac{\lVert P_{Y_t}-P_Z\rVert_1}{\epsilon}\nonumber\\
&\leq\frac{g_{\max}}{\epsilon}\sum_{t\in\setb}P_B(t)\lVert P_{Y_t}-P_Z\rVert_1\nonumber\\
&\oeq{b}\frac{g_{\max}}{\epsilon}\expop\Bigl[\lVert P_{Y_B}-P_Z\rVert_1\Big|P_B\Bigr]\nonumber\\
&\oleq{c}\frac{g_{\max}}{\epsilon}\theta.\label{eq:sum2bound}
\end{align}
Step (a) follow from the assumption in the proposition and we used definition \eqref{eq:conditionaldef} in (b). Inequality (c) follows from the supposition $\expop[\lVert P_{Y_B}-P_Z\rVert_1|P_B]<\theta$. Using the two bounds \eqref{eq:sum1bound} and \eqref{eq:sum2bound} in \eqref{eq:iii:sums}, we get
\begin{align}
\Bigl|\expop\bigl[g(P_{Y_B})\big|P_B\bigr]-g(P_Z)\Bigr|\leq \delta(\epsilon) + \frac{\theta}{\epsilon} g_{\max}.
\end{align}

\subsection{Proof of Prop.~\ref{prop:entropyrate}}
\label{proof:entropyrate}
We apply Prop.~\ref{prop:continuity} with
\begin{align}
g&=\entop\\
g_{\max}&=\log_2|\setz|\\
\delta(\epsilon)&\oeq{a}-\epsilon\log_2\frac{\epsilon}{|\mathcal{Z}|},\quad 0\leq\epsilon\leq\frac{1}{2}
\end{align}
where we apply \cite[Lemma~2.7]{csiszar2011information} in (a). We have
\begin{align}
\left|\frac{\entop(P_Y)}{\expop[\ell(Y)]}-\entop(P_Z)\right|&\oeq{a}\Bigl|\entop(P_{Y_B}|P_B)-\entop(P_Z)\Bigr|\\
&\oleq{b}\delta(\epsilon) + \frac{\theta}{\epsilon} \log_2|\setz|\nonumber\\
&\oeq{c}-\epsilon\log_2\frac{\epsilon}{|\mathcal{Z}|}+\frac{\theta}{\epsilon} \log_2|\mathcal{Z}|\label{eq:predeltaepsilon}.
\end{align}
This bound holds for all $\epsilon$ and we minmize it by calculating its derivative with respect to $\epsilon$ and setting it equal to zero:
\begin{align}
&\hspace{-1cm}\frac{\partial}{\partial\epsilon}\left[-\epsilon\log_2\frac{\epsilon}{|\mathcal{Z}|}+\frac{\theta}{\epsilon} \log_2|\mathcal{Z}|\right]\\
&=-\log_2\frac{\epsilon}{|\setz|}-\log_2 e-\frac{\theta}{\epsilon^2}\log_2|\setz|\overset{!}{=}0\\
\Rightarrow&\theta(\epsilon) := \frac{\epsilon^2}{\log_2|\setz|}\log_2\frac{|\setz|}{e\epsilon}.
\end{align}
We plug $\theta(\epsilon)$ into \eqref{eq:predeltaepsilon} and define
\begin{align}
\sigma(\epsilon):=\epsilon\log_2\frac{|\setz|^2}{e\epsilon^2}.	
\end{align}

\bibliographystyle{IEEEtran}
\normalsize
\bibliography{IEEEabrv,confs-jrnls,references}

\end{document}